\documentclass[10pt]{IEEEtran}
\IEEEoverridecommandlockouts
\usepackage{cite}
\usepackage{amsmath,amssymb,amsfonts}
\usepackage{algorithmic}
\usepackage{tabularx}
\usepackage{adjustbox}
\usepackage{booktabs}
\usepackage{multirow}
\usepackage{array}
\usepackage{makecell}
\usepackage{graphicx}
\usepackage{textcomp}
\usepackage{comment}
\usepackage{xcolor}
\usepackage{cleveref}
\usepackage{braket}
\usepackage{enumitem}
\usepackage{adjustbox}
\usepackage{tikz}
\usepackage{bm}
\usepackage[table]{xcolor} % for \rowcolor
\definecolor{TopK}{RGB}{220,255,220} % light green (adjust if desired)

\usepackage{enumitem}
\setlist[itemize]{leftmargin=*}%
\setlist[enumerate]{leftmargin=*}%
\def\BibTeX{{\rm B\kern-.05em{\sc i\kern-.025em b}\kern-.08em
    T\kern-.1667em\lower.7ex\hbox{E}\kern-.125emX}}

\usepackage{fancyhdr}
\fancypagestyle{firstpage}{
   \fancyhf{} % clear all header and footer fields
   \fancyhead[C]{To appear at 2026 IEEE International Joint Conference on Neural Networks (IJCNN) Proceedings, Maastricht, Netherlands} % Centered header
}

\begin{document}

\title{GAT-QNN: Genetic Algorithm-Based Training of Hybrid Quantum Neural Networks
\vspace{10pt}
}

\author{\IEEEauthorblockN{Tasnim Ahmed\IEEEauthorrefmark{1}\IEEEauthorrefmark{2}, Alberto Marchisio\IEEEauthorrefmark{1}\IEEEauthorrefmark{2}, Muhammad Kashif \IEEEauthorrefmark{1}\IEEEauthorrefmark{2}, Nouhaila Innan\IEEEauthorrefmark{1}\IEEEauthorrefmark{2},
Muhammad Shafique\IEEEauthorrefmark{1}\IEEEauthorrefmark{2}}

\IEEEauthorblockA{\IEEEauthorrefmark{1} \normalsize eBrain Lab, Division of Engineering, New York University Abu Dhabi, PO Box 129188, Abu Dhabi, UAE\\}
\IEEEauthorblockA{\IEEEauthorrefmark{2} \normalsize Center for Quantum and Topological Systems, NYUAD Research
Institute, New York University Abu Dhabi, UAE}

Emails: \{tasnim.ahmed, alberto.marchisio, muhammadkashif, nouhaila.innan, muhammad.shafique\}@nyu.edu

%\vspace{-20pt}
}

\begin{comment}
\author{\IEEEauthorblockN{1\textsuperscript{st} Given Name Surname}
\IEEEauthorblockA{\textit{dept. name of organization (of Aff.)} \\
\textit{name of organization (of Aff.)}\\
City, Country \\
email address or ORCID}
\and
\IEEEauthorblockN{2\textsuperscript{nd} Given Name Surname}
\IEEEauthorblockA{\textit{dept. name of organization (of Aff.)} \\
\textit{name of organization (of Aff.)}\\
City, Country \\
email address or ORCID}
\and
\IEEEauthorblockN{3\textsuperscript{rd} Given Name Surname}
\IEEEauthorblockA{\textit{dept. name of organization (of Aff.)} \\
\textit{name of organization (of Aff.)}\\
City, Country \\
email address or ORCID}
\and
\IEEEauthorblockN{4\textsuperscript{th} Given Name Surname}
\IEEEauthorblockA{\textit{dept. name of organization (of Aff.)} \\
\textit{name of organization (of Aff.)}\\
City, Country \\
email address or ORCID}
\and
\IEEEauthorblockN{5\textsuperscript{th} Given Name Surname}
\IEEEauthorblockA{\textit{dept. name of organization (of Aff.)} \\
\textit{name of organization (of Aff.)}\\
City, Country \\
email address or ORCID}
\and
\IEEEauthorblockN{6\textsuperscript{th} Given Name Surname}
\IEEEauthorblockA{\textit{dept. name of organization (of Aff.)} \\
\textit{name of organization (of Aff.)}\\
City, Country \\
email address or ORCID}
}
\end{comment}

\maketitle
\pagestyle{empty}
\thispagestyle{empty} %page number
\thispagestyle{firstpage}
\begin{abstract}
%Hybrid Quantum Neural Networks (HQNNs) offer promising potential by combining the strengths of classical and quantum processing. However, designing effective quantum circuits for HQNNs is challenging due to the noise in Noisy Intermediate-Scale Quantum (NISQ) devices and the vast design space. To address these challenges, we propose a genetic algorithm framework that trains macroCircuits by sampling and training microCircuits. This approach is beneficial because it yields a set of optimal microcircuits that can be deployed across different devices, eliminating the need for retraining when seeking the best architecture for a specific task and device. By leveraging this large trained design space, our method enhances performance and adaptability across various devices and tasks.

Hybrid Quantum Neural Networks (HQNNs) combine classical learning with parameterized quantum circuits, but their practical performance is often limited by (i) the noise of Noisy Intermediate-Scale Quantum (NISQ) devices and (ii) the large, discrete design space of quantum circuit architectures.
Moreover, HQNNs are commonly trained using a fixed circuit and a single backend, even though deployment frequently targets heterogeneous backends where compilation and execution characteristics may differ.
To address these challenges, we propose \emph{GAT-QNN}, a genetic algorithm (GA)-based framework that trains a \emph{macroCircuit} (search space) by iteratively sampling \emph{microCircuits} (subcircuits), training them, and reintegrating their learned parameters into the macroCircuit.
After training, we run an independent GA-driven inference stage that evaluates candidate microCircuits using the trained macroCircuit weights and selects top-performing architectures for deployment.
This two-stage approach enables backend-aware microCircuit selection without retraining each candidate architecture and can also reduce computational resources (gate count) by deploying smaller microCircuits derived from the macroCircuit.
We validate the approach on MNIST classification (four classes) and report consistent 22-23\% test accuracy gains for GA-driven inference across multiple backends.

\end{abstract}


\section{Introduction}

Quantum Machine Learning (QML) is an emerging paradigm that combines machine learning with quantum computing to address challenges in high-dimensional data processing and complex task analysis~\cite{schuld2015introduction,kashif2025computational,zaman:POQA}. Its potential applications span diverse domains, including finance, healthcare, navigation and computer vision~\cite{innan:quav,zaman2023survey, dutta2024qadqn, innan2025lep, ahmad2026quantum, innan2024quantum,  kashif2021design, innan2025qfnn, innan2025fl, innan2025qnn, siddiqui2025quantum, el2025robqfl, dave2025sentiqnf,dutta2025quiet}. Within QML, Quantum Neural Networks (QNNs) have emerged as a promising class of models that integrate classical neural network principles with quantum computing fundamentals~\cite{kashif2026design}. A QNN architecture typically embeds parameterized quantum circuits (PQCs) as computational layers with trainable components, enabling tasks like quantum-enhanced feature encoding or solving classically intractable problems~\cite{benedetti2019parameterized,kashif_demonstrating}. For example, QNNs can process entangled quantum states or exploit superposition to evaluate multiple solutions simultaneously, offering theoretical speedups for specific tasks~\cite{shor1994algorithms, grover1996fast, kashif2025computational}.

However, the practical implementation of QNNs faces significant barriers due to the limitations of current quantum hardware and noise~\cite{kashif:investigating,kashif:PP}. We are in the Noisy Intermediate-Scale Quantum (NISQ) era, characterized by processors with limited qubit counts (50-100 qubits) and high susceptibility to errors such as decoherence, gate inaccuracies, and readout noise~\cite{kashif2024nrqnn,kashif:hqnet}. These constraints render purely quantum models infeasible for real-world applications~\cite{preskill2018nisq}.

In this regard, Hybrid Quantum Neural Networks (HQNNs) have been developed. HQNNs integrate classical neural layers with shallow PQCs, leveraging quantum expressiveness while maintaining compatibility with NISQ-era constraints~\cite{kashif2023unified}. By combining classical pre-processing with quantum post-processing, HQNNs reduce the number of trainable parameters and enhance computational efficiency. This hybrid design can reduce the number of quantum parameters and keep circuits shallow enough to be compatible with NISQ devices, while still leveraging quantum effects such as entanglement~\cite{mitarai2018quantum}.

\subsection{Target Research Problem and Challenges}
In practice, HQNN circuits are not executed in a single, homogeneous environment.
A model may be trained on a simulator but deployed on heterogeneous backends (e.g., different simulators, compilation flows, or hardware-oriented execution stacks), each introducing backend-specific effects such as compilation-induced structural changes, differences in gate implementations, and distinct noise characteristics~\cite{ahmed2025comparative, ji2025algorithm, ahmed2025noisy}.
Consequently, a circuit architecture that performs well under one backend may not be optimal under another, and backend-agnostic training can lead to brittle deployment behavior~\cite{wang2022quantumnas}.
In addition, deploying a full quantum circuit (macroCircuit) can be unnecessarily expensive, since when a smaller subcircuit (microCircuit) achieves comparable (or better) performance, reducing the gate count is advantageous for both runtime and noise exposure~\cite{preskill2018nisq}.

\subsection{Proposed Approach}
To address these challenges, this paper proposes \textit{GAT-QNN}, a genetic algorithm (GA)-based training and deployment framework for HQNNs that explicitly leverages a large, discrete architectural search space while enabling resource-aware selection.
The key idea is to train a \emph{macroCircuit} that defines a super-structure (search space) and to iteratively sample \emph{microCircuits} (sub-architectures) from it.
During GA-driven training, each sampled microCircuit is trained, and its learned parameters are reintegrated into the macroCircuit, allowing the macroCircuit to accumulate knowledge across many architectural candidates.
After training, we run an \emph{independent} GA-driven inference stage in which candidate microCircuits are evaluated using the trained macroCircuit weights, but without any parameter updates.
This decoupled training/inference design enables selecting microCircuits that best match the target backend at deployment time, while also enabling the selection of lower-cost circuits in terms of gate count.

\subsection{Motivational Case Study}
\label{subsec:motivation}

Table~\ref{tab:motivation_comparison} illustrates why a backend-aware, microCircuit-centric deployment strategy is needed.
For the macroCircuit chromosome $[4,4,4,4,2]$, regular training yields strong accuracy (0.915), yet the GA-trained macroCircuit evaluated directly at the macroCircuit level attains a lower accuracy (0.780).
In contrast, when deployment uses GA-driven inference to select a microCircuit, the chromosome $[3,2,2,1,2]$ achieves substantially higher accuracy under the GA-trained pipeline (0.870) than under regular training (0.320), while also using fewer resources (5 \texttt{RX}, 3 \texttt{CNOT}).
This example highlights two practical takeaways: (i) the GA-trained pipeline is most effective when paired with GA-driven inference-time selection of microCircuits, and (ii) better accuracy can coincide with reduced gate count, which is especially valuable under NISQ constraints.

\begin{table}[t]
\centering
\caption{Motivational example comparing regular training vs.\ GA-trained performance for the macroCircuit (Chromosome [4,4,4,4,2]) and a representative microCircuit architecture (Chromosome [3,2,2,1,2]) architectures, together with their resource usage.}
\label{tab:motivation_comparison}
\setlength{\tabcolsep}{6pt}
\begin{adjustbox}{max width=\linewidth}
\renewcommand{\arraystretch}{1.15}
\begin{tabular}{@{}lcc l@{}}
\toprule
\textbf{Chromosome} & \textbf{Regular Tst Acc} & \textbf{GA-Trained Tst Acc} & \textbf{Resources} \\
\midrule
$[4,4,4,4,2]$   & 0.915 & 0.780 & 8 \texttt{RX}, 8 \texttt{CNOT} \\
$[3,2,2,1,2]$   & 0.320 & \textbf{0.870} & \textbf{5 \texttt{RX}, 3 \texttt{CNOT}} \\
\bottomrule
\end{tabular}
\end{adjustbox}
\end{table}

\subsection{Our Novel Contributions}

\begin{itemize}
    %\item GA-based framework for training HQNNs by co-evolving macrocircuits and microcircuits.
    %\item Demonstrate enhanced robustness and performance on classfication task. 

\item We propose \textit{GAT-QNN}, a GA-based training framework for HQNNs that trains a \emph{macroCircuit} by iteratively sampling, training, and reintegrating \emph{microCircuits}, enabling joint exploration of architectures and parameters under NISQ constraints.

\item We introduce a GA-driven co-evolution process (parent selection, crossover, mutation, and generation-to-generation evolution) that determines which microCircuits are trained/evaluated and how their learned parameters are aggregated into the macroCircuit.

\item We decouple training from deployment by producing a population of optimized microCircuits that can be evaluated and selected for different backends, reducing the need to retrain each candidate architecture when targeting a new backend.

\item We define a two-stage pipeline (GA-driven macroCircuit training followed by an independent GA-based inference/evaluation run) to select microCircuit architectures that maximize inference-time fitness on the target backend.

\item We demonstrate up to 23\% improved performance on an image classification task when using GA-trained macroCircuits together with GA-driven inference-time microCircuit selection, compared to the GA-driven inference-time microCircuit selection on the conventional macroCircuit training, across multiple backends.

\end{itemize}

\section{Background and Related Work}

\subsection{Hybrid Quantum Neural Networks}

\begin{figure}[t]
  \centering
  \includegraphics[width=\linewidth]{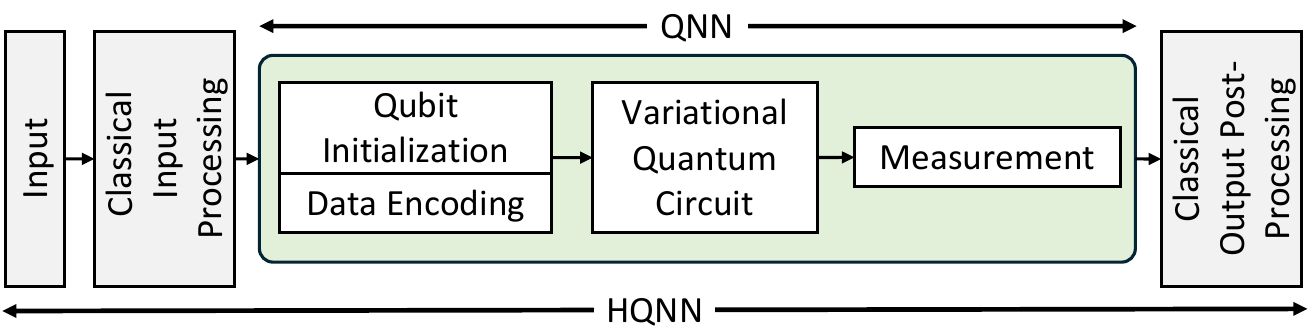}
  \caption{Hybrid quantum neural network (HQNN) pipeline used in this work: classical preprocessing transforms the input into an \(n\)-qubit-compatible feature vector, which is encoded into a quantum state; a shallow parameterized quantum circuit applies trainable single-qubit rotations and entangling operations; expectation-value measurements are returned to the classical domain for post-processing and final prediction.}
  \label{fig:hqnn}
\end{figure}

HQNNs combine classical neural processing with a PQC layer to form an end-to-end trainable model (Fig.~\ref{fig:hqnn}).
The classical component typically handles input conditioning and lightweight post-processing, while the quantum component provides a compact, expressive transformation implemented as a variational circuit\cite{kashif2023impact}.

An HQNN used for supervised learning can be summarized as a composition of four stages:

\begin{enumerate}

    \item \textbf{Classical preprocessing:} The input sample is optionally normalized and transformed (e.g., via feature reduction) into a vector whose dimension matches the quantum encoding scheme.
    This stage may also include classical layers that learn a task-specific embedding before the quantum layer.
    \item \textbf{Classical-to-quantum encoding:} The preprocessed features are loaded into an \(n\)-qubit quantum state using a fixed encoding map (e.g., amplitude or angle encoding), which determines how classical information is represented on the quantum domain.
    \item \textbf{Variational quantum circuit:} A shallow PQC parameterized by a set of trainable rotation angles applies alternating patterns of single-qubit rotations and entangling operations.
    Since the circuit depth is constrained by the NISQ setting, the design of gate patterns and entanglement structure strongly impacts trainability and performance.
    \item \textbf{Measurement and classical post-processing:} One or more expectation values (or sampled bitstrings) are measured from the quantum state and passed to a classical computation stage (e.g., linear layer + softmax) to produce the final prediction.
\end{enumerate}

HQNN parameters include both the classical weights and the PQC rotation angles.
They are optimized jointly using standard supervised objectives (e.g., cross-entropy for classification), where gradients are computed through the quantum layer via differentiable programming interfaces (e.g., parameter-shift) and then propagated to the classical components.
This hybrid training procedure enables integrating quantum layers into conventional deep learning pipelines while respecting shallow-circuit constraints~\cite{innan2025next}.

Since the quantum layer must remain shallow, the placement and number of rotation and entangling gates become critical design choices~\cite{Zaman_2024arxiv_StudyingImpactQuantumHyperparameters, marchisio2025cutting, zaman2024comparative,el2026comparative}.
Over-parameterized circuits can suffer from optimization difficulties and unnecessary resource overhead, while under-parameterized circuits may lack expressiveness.
This motivates automated procedures that explore circuit structures under resource constraints, especially when the deployment backend can influence effective circuit behavior.

\subsection{Genetic Algorithms}

\begin{figure}
  \centering
  \includegraphics[width=\linewidth]{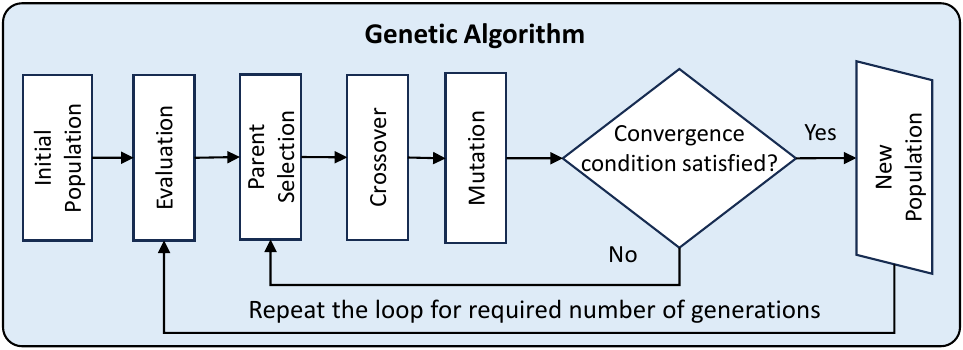}
  \caption{Genetic algorithm workflow adopted in this work. An initial population of chromosomes is evaluated with a task-specific fitness function; the top-performing individuals are selected as parents; multi-point crossover and random mutation generate offspring; the offspring form the next generation, and the loop is repeated for a fixed number of generations (or until convergence).}
  \label{fig:ga}
\end{figure}

A genetic algorithm is an evolutionary search strategy that iteratively improves a population of candidate solutions by favoring high-fitness individuals (Fig.~\ref{fig:ga}).
It is particularly suitable for discrete and combinatorial search spaces, such as quantum circuit structure selection, where gradient-based optimization is not directly applicable to architectural choices.

%Genetic Algorithm (GA) is an evolutionary search method inspired by Darwin’s theory of natural selection, mirroring biological evolution in both name and function. At each step, the algorithm selects the fittest individuals as parents, combines and mutates them to produce offspring, and gradually evolves the population toward an optimal solution. GAs are widely used for complex optimization problems, especially in machine learning for tasks like hyperparameter tuning, feature selection, and model architecture optimization, where traditional methods may struggle with large or non-linear search spaces.

GAs are composed of the following core components:

\begin{enumerate}
    \item \textbf{Population and chromosome:} In a genetic algorithm, the population is a set of candidate solutions, each called an individual. Each individual is represented by a chromosome, with each parameter in the chromosome corresponding to a gene. Chromosomes can be encoded as binary strings, arrays, or other data structures. Each gene encodes a discrete design decision (e.g., circuit depth or per-layer gate counts). %The population then undergoes an iterative process in which individuals are selected based on their fitness, guiding the search toward optimal solutions.
    \item \textbf{Population and chromosome:} Each chromosome is assigned a fitness score, computed by an objective function (e.g., validation/test accuracy for ML classification tasks), which evaluates how well the individual solves the problem at hand. The fitness function drives the evolutionary process by ranking individuals and guiding the algorithm toward better solutions.
    \item \textbf{Parent Selection:} 
    A subset of high-fitness candidates is selected as parents, based on their rank compared to other candidates. In practice, in each generation, individuals with higher fitness values are more likely to be selected, mimicking the natural survival selection. In this way, the best solutions are preserved and they guide the process across generations. 
    \item \textbf{Crossover and Mutation:} New individuals are generated during the evolutionary process through \textit{crossover} operations, which recombine the traits of two parents, with the aim of producing offspring that inherit beneficial characteristics from both parents. Following crossover, random \textit{mutations} (e.g., flipping a bit or changing a gene in the chromosome structure) are applied to some individuals, introducing slight variations that help maintain diversity within the population and explore new areas of the solution space.
\end{enumerate}

In this paper, a GA provides a natural mechanism to explore discrete microCircuit configurations derived from a macroCircuit search space.
By repeatedly evaluating, selecting, and evolving candidate chromosomes, the method can identify architectures that achieve strong task performance while implicitly accounting for resource usage and backend-dependent effects.

\subsection{Related Works}
%- Discuss the neural architecture search process: specifics of micro and macro - 
%- other GA related paper in qml training

\begin{figure*}[h]
    \centering
    \includegraphics[width=\linewidth]{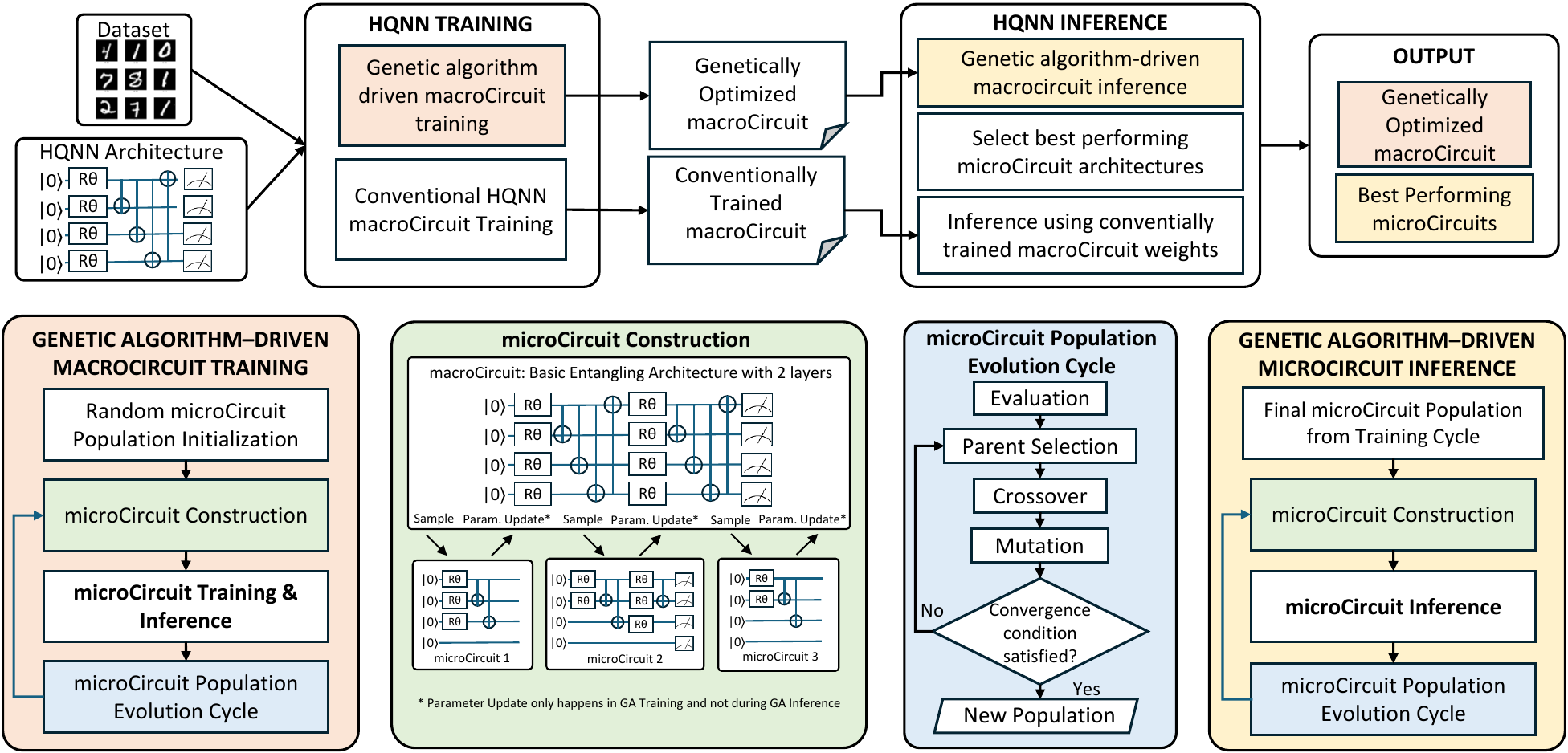}
    \caption{\footnotesize Overview of the proposed GAT-QNN methodology. The macroCircuit defines a discrete architectural search space over layered RX (parameterized single-qubit rotations) and CNOT (entangling) gates. \textbf{Training stage (GA-driven macroCircuit training):} a genetic algorithm evolves a population of chromosomes, where each chromosome instantiates a microCircuit (subcircuit) sampled from the macroCircuit; each sampled microCircuit is trained on the task, evaluated with test accuracy as fitness, and its learned parameters are reintegrated into the shared macroCircuit weights (parameter update occurs only in this stage). \textbf{Deployment stage (GA-driven macroCircuit inference):} starting from the final training population, an independent GA run performs inference-only evaluation of candidate microCircuits using the trained macroCircuit weights (no parameter updates), and returns a ranked set of microCircuits. The final output is the set of best-performing microCircuit architectures selected for deployment on the target backend, enabling backend-aware selection and reduced gate count compared to executing the full macroCircuit.}
    \label{fig:methodology}
\end{figure*}

Recent works have explored aspects which are related to the proposed GAT-QNN framework, such as (i) quantum architecture search (QAS) for parameterized quantum circuits, (ii) evolutionary and multi-objective search for QNN/HQNN architectures, and (iii) training approaches that amortize the cost of evaluating many candidate circuits.
A central limitation across much of the prior literature is that it often returns a \emph{single} optimized architecture for a particular simulator/noise setting, without a clear mechanism to \emph{train once} and then \emph{adapt deployment} by selecting a suitable microCircuit on the target backend without retraining.

\textbf{Variational training challenges.}
Several widely recognized phenomena motivate resource-aware and robust QNN design.
Barren plateaus can cause vanishing gradients in variational circuits, complicating training and gradient propagation as circuits scale~\cite{mcclean2018barren,kashif:alleviating,kashif2024resqnets,kashif2025deep}
Foundational works on variational quantum algorithms in the NISQ setting highlight why shallow circuits and careful architectural choices are required in practice~\cite{cerezo2021vqa,preskill2018nisq}. %,kashif2023impact.
These challenges strengthen the need for frameworks that (i) avoid repeatedly training large circuits from scratch and (ii) enable selecting smaller, well-performing quantum circuits for practical deployment.

\textbf{QAS for variational circuits (efficiency vs.\ deployment).}
Weight-sharing QAS methods reduce evaluation cost by reusing parameters across many candidate circuits, enabling practical exploration of large discrete spaces.
Du et al. propose a quantum circuit architecture search method for variational quantum algorithms using weight sharing, showing that one-shot style training can accelerate architecture exploration~\cite{du2022qas}.
Differentiable search further reduces reliance on sampling-based evaluation: QuantumDARTS uses a differentiable relaxation (Gumbel-Softmax) to search PQC architectures for several tasks~\cite{wu2023quantumdarts}.
While these methods improve search efficiency, they typically focus on selecting an architecture within a fixed execution assumption and do not explicitly decouple training from inference-time architecture selection on heterogeneous backends.

\textbf{Noise- and hardware-aware QAS.}
Noise-aware search emphasizes that circuit structure and hardware constraints fundamentally shape achievable performance in the NISQ regime.
QuantumNAS performs noise-adaptive search for robust quantum circuits, jointly considering architectural choices and hardware effects, and uses a SuperCircuit weight-sharing mechanism to amortize training~\cite{wang2022quantumnas}.
Similarly, He et al. accelerate QAS by combining Bayesian optimization with a self-supervised surrogate that learns circuit representations to reduce expensive architecture evaluations~\cite{he2025ssrlqAS}.
Recent work also explores multi-objective, noise-aware formulations; for example, the NSGA-II-based NA-QAS trades off performance and hardware overhead under a noise model~\cite{li2026naqas}.
These approaches are aiming at obtaining a robust design, but they still largely optimize toward a final architecture (or Pareto set) rather than explicitly producing a trained macroCircuit that supports \emph{inference-time} GA selection of microCircuits on the target backend without parameter updates.

\textbf{Evolutionary HQNN/QNN NAS and resource-awareness.}
Evolutionary methods are well-suited for discrete gate/topology decisions and have been applied to QNN design.
EQNAS proposes an evolutionary quantum neural architecture search method for image classification and reports improvements in accuracy and reduced parameter and gate counts relative to baselines~\cite{li2023eqnas}.
FAQNAS formulates HQNN design as a multi-objective search balancing accuracy and FLOPs, using the NSGA-II genetic algorithm~\cite{kashif2025faqnas}.
These works reinforce that GA-based methods can discover strong architectures and manage resource trade-offs; however, FLOPs- or gate-awareness alone does not address the training-deployment mismatch across heterogeneous backends, and these approaches typically do not define a two-stage pipeline where (i) a shared macroCircuit is trained, then (ii) an independent GA run selects microCircuits at inference time for the target backend.

\textbf{Positioning of GAT-QNN.}
In contrast to prior work that primarily outputs a best-found architecture for a given assumption, GAT-QNN explicitly trains a macroCircuit by iteratively sampling/training microCircuits and reintegrating their parameters, then performs an independent GA-driven inference stage to select microCircuits on the target backend without parameter updates.
This directly targets heterogeneous-backend robustness and resource-aware deployment through microCircuit selection, complementing weight-sharing QAS, differentiable QAS, and evolutionary multi-objective HQNN NAS~\cite{du2022qas,wu2023quantumdarts,wang2022quantumnas,li2023eqnas,kashif2025faqnas,li2026naqas, kashif2026closing, maleki2026qnas,dutta2025qas,innan2025circuithunt,choudhary2025graph}.

\section{GAT-QNN Methodology}

Figure~\ref{fig:methodology} summarizes the proposed end-to-end pipeline, which consists of (i) GA-driven macroCircuit training, (ii) GA-driven macroCircuit inference, and (iii) microCircuit selection based on inference-time fitness, i.e., test accuracy.
The macroCircuit acts as a super-structure that defines a large architectural search space. At each iteration, the genetic algorithm samples microCircuits (sub-structures) whose measured accuracy provides an on-backend estimate of architectural quality.

%\subsection{Conventional MacroCircuit Training}
%- trained basic entangling circuit regularly, on gpu 
%- pennylane - lightning qubit, training sample: 8000, testing sample: 2000, batch size: 20, epochs: 20
%- training hqnn process : pca downsampling, input into 4 qubit basic entangling circuit, circuit measured, circuit gates trained with classical optimization method (adam optimizer, lr = 0.01)

\subsection{Chromosome Encoding}
\label{sec:chromosome_encoding}

We encode each candidate microcircuit architecture as a fixed-length integer chromosome, following the regular convention used in recent macroCircuit-based search frameworks, where a large macroCircuit implicitly defines a search space of microCircuits, and the search algorithm selects discrete architectural choices within that space.

\textbf{Gene structure:}
A chromosome is represented as
\begin{equation}
\mathbf{g} = \big[g^{(1)}_{R},\; g^{(1)}_{C},\; g^{(2)}_{R},\; g^{(2)}_{C},\; L \big],
\end{equation}
where \(L\) denotes the number of active layers, and the remaining genes control the width of each layer.
Specifically, for layer \(\ell \in \{1,\dots,L\}\), \(g^{(\ell)}_{R}\) specifies the number of parameterized single-qubit rotation gates (implemented as \texttt{RX}), and \(g^{(\ell)}_{C}\) specifies the number of entangling gates (implemented as \texttt{CNOT}). 
Only the first \(2L\) genes are considered active; genes beyond layer \(L\) are ignored during circuit construction.

\textbf{Value ranges and search space:}
For the 4-qubit experiments, each rotation-width and entanglement-width gene is drawn from \(\{1,2,3,4\}\), while the depth gene is drawn from \(\{1,2\}\), yielding a compact yet expressive discrete design space.
This encoding can be extended to larger qubit counts by expanding the admissible gene values accordingly.

\textbf{Mapping from chromosome to circuit:}
Given a chromosome \(\mathbf{g}\) and number of qubits \(n\), the circuit is instantiated by stacking \(L\) layers.
At layer \(\ell\), we apply \(g^{(\ell)}_{R}\) \texttt{RX} gates on the first \(g^{(\ell)}_{R}\) wires, each with an independent trainable parameter.
We then apply \(g^{(\ell)}_{C}\) \texttt{CNOT} gates using a ring-like connectivity pattern among the \(n\) wires, thereby controlling the entanglement strength in a discrete manner.
The total number of trainable parameters for the resulting microcircuit is
\begin{equation}
P(\mathbf{g}) = \sum_{\ell=1}^{L} g^{(\ell)}_{R},
\end{equation}
since each selected \texttt{RX} contributes one rotation angle.

\textbf{Genetic operators:}
This integer-based encoding supports standard genetic operators.
Crossover exchanges contiguous gene segments between two parent chromosomes (multi-point crossover), while mutation perturbs individual genes by resampling them from their admissible sets with a predefined probability.
Because the last gene \(L\) directly controls which layer genes are active, mutation of \(L\) changes circuit depth, whereas mutations of the other genes change per-layer expressivity.

\subsection{MacroCircuit Training with Genetic Algorithm}

\label{subsec:ga_training}

We train the macroCircuit using a genetic algorithm (GA) that evolves a population of microCircuit chromosomes.
Each chromosome encodes a microCircuit instantiated from the macroCircuit search space (Fig.~\ref{fig:methodology}), and its fitness is computed after training and evaluation on the target backend.

\textbf{GA hyperparameters:}
We use the following GA configuration: population size~\(=10\), number of generations~\(=5\), parent fraction~\(=0.4\) (i.e., 4 parents per generation), mutation rate~\(=0.6\), and number of crossover points~\(=3\).

\textbf{Iterative training cycle:}
The GA training cycle proceeds as follows:
\begin{enumerate}
    \item \textit{Initialization:} Generate an initial population of random chromosomes (microCircuit architectures).
    \item \textit{Fitness evaluation with training:} For each chromosome, instantiate the corresponding microCircuit, train it, and compute its fitness (i.e., test accuracy). The macroCircuit is updated by integrating the trained parameters obtained from the evaluated microCircuits, forming a progressively improved parameter pool for the overall search space.
    \item \textit{Parent selection:} Select the top-performing chromosomes as parents using partial elitism (top 4 chromosomes in our setup).
    \item \textit{Crossover and mutation:} Generate candidate offspring via multi-point crossover of selected parents, and apply mutation to introduce diversity and to produce additional chromosomes when needed to restore the target population size.
    \item \textit{Next generation:} The resulting set of chromosomes forms the population for the next generation.
\end{enumerate}
This cycle is repeated for the predefined number of generations, producing a genetically optimized and trained macroCircuit at the end of training.

% as described in section "background". Training with GA entails selecting a certain number of generations you want to repeat the algorithm. In our framework we had the following GA parameters defined: 
% - Population Size: 10
% - Number of Generations: 5
% - Parent Size: 0.4*prevpopulation = 4 chromosomes
% - Mutation Rate: 0.6
% - Crossover Points: 3

% \emph{Iterative Training pipeline}
% \begin{enumerate}
%     \item Pool of random chromosomes (microCircuit architectures) generated in the first generation
%     \item then in the fitness functions each individual chromosomes is trained and the macrocircuit is updated with the trained weight. We store the results of each of the chromosomes with its corresponding training result
%     \item then for parent selection the top performing chromosomes are selected, in our case the 0.4*prevpopulation = top 4 gene
%     \item offspring generation: we perform crossover with the selected parents chromosomes by splitting then at 3 points and combining them and then for remaining chromosomes (we need to match the total population size) we generate them through mutation
%     \item The offsprings become the next generation of initial chromosome pool
% \end{enumerate}

% steps 2-5 repeated for a 10 generations. The final generation gives us the fully trained macroCircuit using genetic algorithm

\begin{table*}[t]
\centering
\caption{Training phase GA results (test accuracy) across generations. Each entry reports a chromosome (microCircuit configuration) and its corresponding test accuracy.}
\label{tab:ga_training_testacc}
\setlength{\tabcolsep}{6pt}
\begin{adjustbox}{max width=\linewidth}
\renewcommand{\arraystretch}{1.15}
\begin{tabularx}{\textwidth}{@{}>{\raggedright\arraybackslash}X r >{\raggedright\arraybackslash}X r >{\raggedright\arraybackslash}X r >{\raggedright\arraybackslash}X r >{\raggedright\arraybackslash}X r@{}}
\toprule
\multicolumn{2}{c}{Generation 1} &
\multicolumn{2}{c}{Generation 2} &
\multicolumn{2}{c}{Generation 3} &
\multicolumn{2}{c}{Generation 4} &
\multicolumn{2}{c}{Generation 5} \\
\cmidrule(lr){1-2}\cmidrule(lr){3-4}\cmidrule(lr){5-6}\cmidrule(lr){7-8}\cmidrule(lr){9-10}
Chromosome & Test Acc &
Chromosome & Test Acc &
Chromosome & Test Acc &
Chromosome & Test Acc &
Chromosome & Test Acc \\
\midrule
$[1,1,4,1,1]$ & 0.817 & $[1,1,4,1,1]$ & 0.820 & $[1,1,4,1,1]$ & 0.820 & $[1,2,4,2,1]$ & 0.822 & $[2,4,1,4,2]$ & 0.882 \\
$[2,4,1,4,1]$ & 0.685 & $[1,1,1,1,2]$ & 0.809 & $[1,2,4,2,1]$ & 0.822 & $[1,2,4,4,1]$ & 0.820 & $[2,4,1,2,2]$ & 0.885 \\
$[3,3,1,4,1]$ & 0.686 & $[2,4,4,4,1]$ & 0.688 & $[2,2,1,4,1]$ & 0.822 & $[2,2,1,4,1]$ & 0.819 & $[2,2,1,3,2]$ & 0.692 \\
$[3,1,3,4,1]$ & 0.820 & $[1,1,1,4,2]$ & 0.656 & $[2,3,4,1,1]$ & 0.684 & $[2,4,1,4,2]$ & 0.883 & $[2,3,1,2,2]$ & 0.703 \\
$[3,4,3,4,1]$ & 0.687 & $[2,4,1,4,2]$ & 0.884 & $[2,4,1,4,2]$ & 0.880 & $[3,2,3,4,1]$ & 0.826 & $[1,4,4,2,2]$ & 0.892 \\
$[2,4,1,4,2]$ & 0.879 & $[3,1,3,4,1]$ & 0.815 & $[3,2,3,4,1]$ & 0.822 & $[3,4,3,4,1]$ & 0.697 & $[1,4,4,4,2]$ & 0.883 \\
$[4,1,3,4,1]$ & 0.814 & $[3,2,3,4,1]$ & 0.823 & $[3,1,3,4,1]$ & 0.813 & $[4,2,3,2,1]$ & 0.825 & $[3,3,2,2,2]$ & 0.759 \\
$[4,2,3,2,1]$ & 0.817 & $[3,1,4,4,1]$ & 0.813 & $[4,2,3,2,1]$ & 0.825 & $[1,4,4,2,2]$ & 0.893 & $[3,4,3,4,2]$ & 0.914 \\
$[3,3,2,2,2]$ & 0.696 & $[4,2,3,2,1]$ & 0.823 & $[4,1,3,1,1]$ & 0.813 & $[3,4,3,4,2]$ & 0.873 & $[4,4,3,2,2]$ & 0.897 \\
$[3,2,4,1,2]$ & 0.817 & $[4,1,3,2,1]$ & 0.814 & $[3,4,3,4,2]$ & 0.863 & $[4,4,3,2,2]$ & 0.911 & $[4,4,3,1,2]$ & 0.835 \\
\bottomrule
\end{tabularx}
\end{adjustbox}
\end{table*}

\subsection{MacroCircuit Inference}
\label{subsec:macro_inference}

After training, we evaluate the trained macroCircuit using two inference modes shown in Fig.~\ref{fig:methodology}: conventional inference and GA-driven inference.
Both modes use the same trained macroCircuit weight state, but differ in whether evolutionary selection is performed at inference time.

\subsubsection{Conventional macroCircuit inference}
\label{subsubsec:conv_inference}

In the conventional baseline, inference is performed directly using the conventionally trained macroCircuit weights, without evolving or selecting microCircuits at inference time.
This represents the standard HQNN evaluation path where architecture is fixed during inference.

\subsubsection{GA-driven macroCircuit inference}
\label{subsubsec:ga_inference}

In GA-driven inference, a second evolutionary process is executed, but the fitness function is inference-only: candidate microCircuits are constructed and evaluated using the already-trained macroCircuit weights, without any parameter updates.
The initial population for GA-driven inference is seeded from the final-generation chromosomes obtained during GA training, ensuring that inference starts from architectures already optimized during training.
This inference-time evolutionary search explicitly accounts for backend-dependent effects by selecting architectures that yield the best measured inference fitness on the target backend.

% \textbf{Conventionally trained circuit:}
% inference the trained macrocircuit on chosen simulators

% \textbf{Genetical Algorithm trained circuit:}
% Inference the trained macrocircuit on chosen simulators

% The steps are same as the Iterative Training pipeline but in the first geenration of genes we use the last generation of chromosomes from the training process and in the fitness function we perform inference instead of training, use the weights from the trained macrocircuit.

\subsection{microCircuit Selection, Inference, and Evaluation}
\label{subsec:micro_selection}

At the end of GA-driven inference, the final population contains a set of candidate microCircuits along with their inference-time fitness values.
We select the top-\(K\) microCircuits (e.g., the best three candidates in Fig.~\ref{fig:methodology}) according to their inference-time fitness.
These selected microCircuits are then reported as the final deployed architectures for inference, and their performance constitutes the output of the proposed methodology.

%from the last generation of GA inference, check which ones perform the best. 

\begin{table*}[t]
\centering
\caption{Inference phase GA results (test accuracy) across generations. Each entry reports a chromosome (microCircuit configuration) and its corresponding test accuracy (top 4 architectures highlighted in bold).}
\label{tab:ga_inference_testacc}
\setlength{\tabcolsep}{6pt}
\begin{adjustbox}{max width=\linewidth}
\renewcommand{\arraystretch}{1.15}
\begin{tabularx}{\textwidth}{@{}>{\raggedright\arraybackslash}X r >{\raggedright\arraybackslash}X r >{\raggedright\arraybackslash}X r >{\raggedright\arraybackslash}X r >{\raggedright\arraybackslash}X r@{}}
\toprule
\multicolumn{2}{c}{Generation 1} &
\multicolumn{2}{c}{Generation 2} &
\multicolumn{2}{c}{Generation 3} &
\multicolumn{2}{c}{Generation 4} &
\multicolumn{2}{c}{Generation 5} \\
\cmidrule(lr){1-2}\cmidrule(lr){3-4}\cmidrule(lr){5-6}\cmidrule(lr){7-8}\cmidrule(lr){9-10}
Chromosome & Test Acc &
Chromosome & Test Acc &
Chromosome & Test Acc &
Chromosome & Test Acc &
Chromosome & Test Acc \\
\midrule
$[2,4,1,4,2]$ & 0.12 & $[1,3,1,1,2]$ & 0.265 & $[3,1,2,1,2]$ & 0.805 & $[2,2,1,2,2]$ & 0.565 & \bm{$[3,2,2,1,2]$} & \textbf{0.87} \\
$[2,4,1,2,2]$ & 0.36 & $[1,4,4,4,2]$ & 0.555 & $[3,2,2,4,2]$ & 0.64 & $[3,4,1,1,2]$ & 0.655 & \bm{$[4,4,1,1,2]$} & \textbf{0.825} \\
$[2,2,1,3,2]$ & 0.475 & $[3,3,2,2,2]$ & 0.535 & $[3,2,2,1,2]$ & 0.865 & $[3,2,2,1,2]$ & 0.87 & \bm{$[4,4,1,2,2]$} & \textbf{0.83} \\
$[2,3,1,2,2]$ & 0.415 & $[1,3,4,4,2]$ & 0.54 & $[3,1,2,4,2]$ & 0.515 & $[4,4,1,1,2]$ & 0.825 & $[3,1,2,1,2]$ & 0.81 \\
$[1,4,4,2,2]$ & 0.53 & $[3,4,2,2,2]$ & 0.6 & $[3,1,2,3,2]$ & 0.59 & $[4,2,2,1,2]$ & 0.27 & $[3,2,3,1,2]$ & 0.815 \\
$[1,4,4,4,2]$ & 0.535 & $[3,1,2,1,2]$ & 0.815 & $[4,4,1,1,2]$ & 0.825 & $[3,3,3,2,2]$ & 0.455 & $[4,4,2,1,2]$ & 0.815 \\
$[3,3,2,2,2]$ & 0.535 & $[3,2,2,4,2]$ & 0.64 & $[4,4,3,1,2]$ & 0.825 & $[4,4,3,1,2]$ & 0.825 & $[4,4,2,2,2]$ & 0.815 \\
$[3,4,3,4,2]$ & 0.22 & $[4,4,3,2,2]$ & 0.815 & $[4,4,3,2,2]$ & 0.815 & $[4,4,3,2,2]$ & 0.815 & \bm{$[4,4,3,1,2]$} & \textbf{0.825} \\
$[4,4,3,2,2]$ & 0.815 & $[4,4,3,1,2]$ & 0.825 & $[3,1,4,1,2]$ & 0.51 & $[4,4,3,4,2]$ & 0.785 & $[4,4,3,2,2]$ & 0.815 \\
$[4,4,3,1,2]$ & 0.815 & $[4,3,3,1,2]$ & 0.22 & $[3,2,4,1,2]$ & 0.81 & $[3,1,4,4,2]$ & 0.545 & $[3,2,4,4,2]$ & 0.6 \\
\bottomrule
\end{tabularx}
\end{adjustbox}
\end{table*}

\section{Results and Discussion}
\subsection{Experimental Setup}

\textbf{Task and dataset:}
We evaluate the proposed approach on a supervised image classification task using MNIST, restricting the dataset to a user-specified subset of 4 classes, i.e., $[0,1,2,3]$. For each selected class, we construct balanced training and test subsets by taking $6\,000$ training samples and $1\,000$ test samples per class.

Note that this is a widely recognized benchmark for QML research, as the constraints of current NISQ devices limit the implementation of HQNNs to small-scale quantum circuits.

\textbf{Input preparation for quantum encoding:}
Each MNIST image is flattened into a 784-dimensional vector and then reduced via PCA to \(2^{n}\) features, where \(n\) denotes the number of qubits.
The resulting feature vectors are \(\ell_2\)-normalized so that each sample can be directly used for amplitude encoding.

\textbf{Hybrid quantum layer implementation:}
Quantum computation is implemented with PennyLane and integrated into a PyTorch training pipeline.
Classical inputs are loaded into the quantum circuit using amplitude embedding over \(n\) qubits, and the quantum layer outputs one expectation value per qubit (Pauli-\(Z\) measurements), which is then consumed by the classical post-processing stage.

\textbf{Baseline macroCircuit architecture:}
As the baseline search space (macroCircuit), we adopt a 4-qubit \emph{basic entangling architecture} composed of up to two stacked layers, where each layer is parameterized by (i) the number of single-qubit rotation operations and (ii) the number of entangling CNOT operations. 
In the GA construction, each of these per-layer gene values is chosen from \(\{1,2,3,4\}\), and the number of active layers is chosen from \(\{1,2\}\), forming the macroCircuit gene template used by the evolutionary search.

\subsection{MacroCircuit Training Evaluation}
\label{subsec:results_macro_training}

Table~\ref{tab:ga_training_testacc} reports the per-generation test accuracy of the GA-evolved microCircuit population during the macroCircuit training stage.
Across generations, the population increasingly concentrates around higher-performing chromosomes, indicating that selection, crossover, and mutation effectively bias the search toward architectures that train well within the macroCircuit parameter-sharing setting.
Notably, the best-performing candidates in later generations achieve test accuracies above 0.90 (e.g., 0.914 in Generation 5), suggesting that the GA can discover highly accurate microCircuit configurations within a small number of generations.

% \subsection{MacroCircuit Inference Evaluation}
% \label{subsec:results_macro_inference}

% Table~\ref{tab:inference_macro_part1} evaluates the macroCircuit configuration (chromosome $[4,4,4,4,2]$) across three backends: PennyLane, AWS Braket simulator, and a QASM simulator.
% The regular (conventional) training baseline achieves consistent accuracy across the three backends (0.915), showing that the evaluated simulators produce matching outcomes for this configuration in our setup.
% In contrast, the GA-trained macroCircuit yields a lower accuracy (0.780) for the same chromosome across all backends. This is due to the fact that chromosome $[4,4,4,4,2]$ only represents one of the possible choices in the design space. As it will be shown in the subsequent microCircuit-level inference stage, the inference-time evolutionary selection identifies the best-performing sub-architectures under the GA-trained weight state, leading to a set of microCircuits that achieve better accuracy than the macroCircuit.

% \begin{table}[t]
% \centering
% \caption{MacroCircuit (chromosome $[4,4,4,4,2]$) test accuracy evaluation across different backends.}
% \label{tab:inference_macro_part1}
% \setlength{\tabcolsep}{6pt}
% \renewcommand{\arraystretch}{1.15}
% \begin{tabular}{@{}lccc@{}}
% \toprule
% \textbf{Training Method} & \textbf{PennyLane} & \textbf{AWS} & \textbf{QASM} \\
% \midrule
% Regular Training & 0.915 & 0.915 & 0.915 \\
% GA Training      & 0.780 & 0.780 & 0.780 \\
% \bottomrule
% \end{tabular}
% \end{table}

\subsection{MicroCircuit Inference Evaluation}

Table~\ref{tab:ga_inference_testacc} reports the inference-time fitness (test accuracy) of the GA-driven microCircuit population and highlights the top-performing architectures selected from the final inference population.

\begin{table}[t]
\centering
\caption{Test accuracy and resource count evaluation of GA-driven microCircuit inference at the Generation 5 (top 4 architectures highlighted).}
\label{tab:inference_micro_part21}
\setlength{\tabcolsep}{6pt}
\renewcommand{\arraystretch}{1.15}
\begin{tabular}{@{}lcl@{}}
\toprule
\textbf{Chromosome} & \textbf{Tst Acc} & \textbf{Resources} \\
\midrule
\rowcolor{TopK} $[3,2,2,1,2]$ & 0.870 & 5 \texttt{RX}, 3 \texttt{CNOT} \\
\rowcolor{TopK} $[4,4,1,1,2]$ & 0.825 & 5 \texttt{RX}, 5 \texttt{CNOT} \\
\rowcolor{TopK} $[4,4,1,2,2]$ & 0.830 & 5 \texttt{RX}, 6 \texttt{CNOT} \\
$[3,1,2,1,2]$                & 0.810 & 5 \texttt{RX}, 2 \texttt{CNOT} \\
$[3,2,3,1,2]$                & 0.815 & 6 \texttt{RX}, 3 \texttt{CNOT} \\
$[4,4,2,1,2]$                & 0.815 & 6 \texttt{RX}, 5 \texttt{CNOT} \\
$[4,4,2,2,2]$                & 0.815 & 6 \texttt{RX}, 6 \texttt{CNOT} \\
\rowcolor{TopK} $[4,4,3,1,2]$ & 0.825 & 7 \texttt{RX}, 5 \texttt{CNOT} \\
$[4,4,3,2,2]$                & 0.815 & 7 \texttt{RX}, 6 \texttt{CNOT} \\
$[3,2,4,4,2]$                & 0.600 & 7 \texttt{RX}, 6 \texttt{CNOT} \\
\bottomrule
\end{tabular}
\end{table}

Note that inference-time test accuracies can be lower than the per-generation training-stage accuracies because, during GA-driven macroCircuit training, each microCircuit is evaluated immediately after its own weight updates and these updated parameters are then aggregated into the macroCircuit at the end of each generation, whereas during GA-driven inference the microCircuits are evaluated in a strictly inference-only manner using the fixed macroCircuit weights obtained after the final (Generation~5) aggregation, with no further parameter updates.

\begin{figure}[t]
    \centering
    \includegraphics[width=\linewidth]{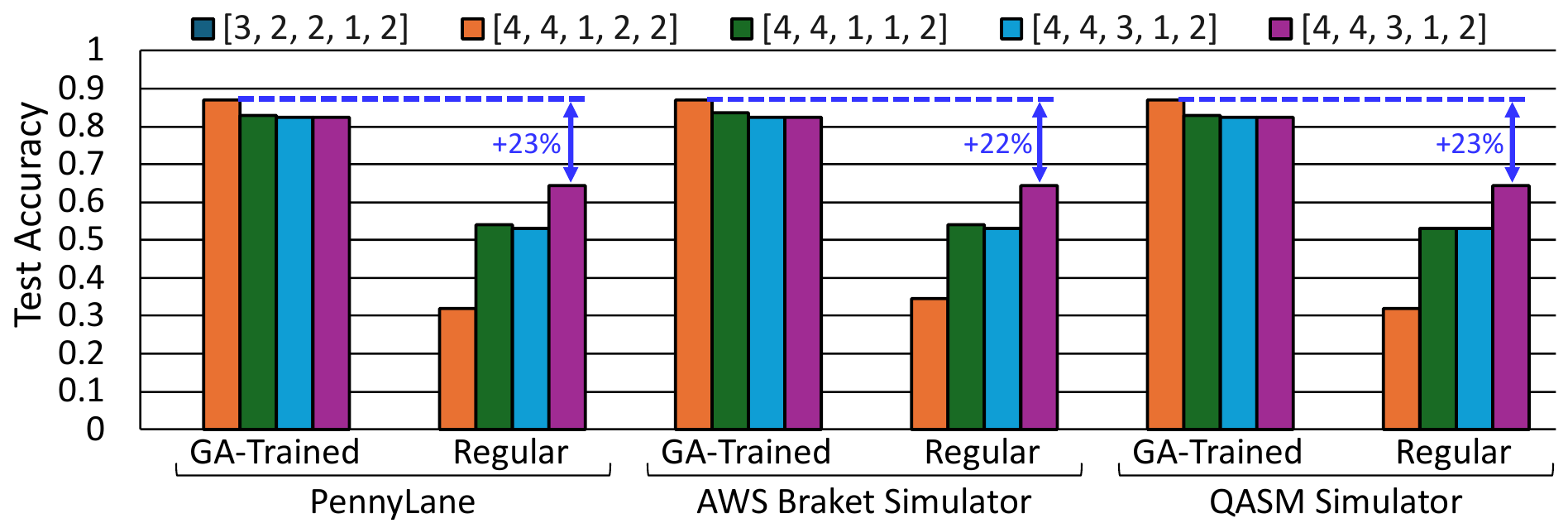}
    \caption{\footnotesize Cross-backend test accuracy comparing GA-trained versus regularly trained macroCircuits under GA-driven inference-time microCircuit selection. Results are reported for three execution backends (PennyLane, AWS Braket simulator, and a QASM simulator) on the MNIST dataset, showing that the GA-trained pipeline consistently achieves higher test accuracy than the regular baseline across all backends.}
    \label{fig:results_simulators}
\end{figure}

The best microCircuit candidate achieves 0.870 test accuracy (chromosome $[3,2,2,1,2]$), while multiple competitive candidates cluster around 0.825--0.830, indicating that the inference stage surfaces several accurate deployment options for this task.
These results support the workflow shown in Fig.~\ref{fig:results_simulators}: when inference is performed via the GA-driven inference procedure, the GA-trained variant consistently achieves higher test accuracy than the regular baseline.
In addition, the performance trend is consistent across all evaluated backends (+23\% on PennyLane, +22\% on AWS Braket simulator, and +23\% on the QASM simulator), indicating that the observed improvement is not specific to a single simulator implementation.
This cross-backend consistency strengthens the claim that GA-driven training and inference yields a more robust deployment strategy under heterogeneous execution backends.

\subsection{Key Observations and Discussion}

As shown in Table~\ref{tab:ga_training_testacc}, across training generations, the GA-evolved population concentrates on higher-fitness chromosomes, indicating that selection and variation operators successfully bias the search toward microCircuit configurations that can be trained effectively within the shared macroCircuit parameter pool.
In addition, the inference-time GA stage surfaces multiple competitive deployment candidates (Table~\ref{tab:inference_micro_part21}), with the best microCircuit achieving strong accuracy while using fewer gates than the full macroCircuit, demonstrating resource-aware deployment under NISQ constraints.
These observations show that robustness is not only a function of training quality but also of \emph{deployment-time} architectural choice. By decoupling training (parameter accumulation in the macroCircuit) from inference-time selection (fitness-driven microCircuit choice), the GAT-QNN framework provides a practical mechanism to mitigate backend-dependent effects without retraining each candidate architecture.

A second fundamental observation is that the GA-trained macroCircuit is most beneficial when paired with GA-driven inference, because the trained macroCircuit acts as a reusable parameter model whose utility is realized through selecting an appropriate microCircuit for the target backend.
This helps explain why direct macroCircuit evaluation is not necessarily the best indicator of deployment performance. The macroCircuit’s role is to define and train a \emph{design space}, whereas the inference-time GA selects microCircuits that best trade off performance and hardware-relevant resource usage for the given execution context.
These results validate and support the methodological choice of a two-stage pipeline (GA-driven macroCircuit training followed by an independent GA-driven inference run) as a robust deployment strategy that can simultaneously achieve high accuracy and reduce gate count compared to executing the full circuit.

\section{Conclusion}

This paper introduced GAT-QNN, a GA-based framework for training and deploying HQNNs that trains a macroCircuit by iteratively sampling, training, and reintegrating microCircuits, followed by an independent GA-driven inference stage that selects top-performing microCircuits on the target backend without parameter updates.
Experiments on a multi-class MNIST classification task show that the GA training process identifies high-performing microCircuit configurations and that inference-time GA selection yields a set of accurate, resource-efficient deployment candidates, consistently outperforming by 22-23\% the microCircuits extracted from the regularly-trained model, across various backends.
These findings motivate future work on scaling the macro/micro search space to larger qubit counts, extending fitness objectives to explicitly incorporate hardware-aware cost models and device-calibrated error rates, and validating the approach on real quantum hardware and broader benchmark suites.

\section*{Acknowledgments} 

This work was supported in part by the NYUAD Center for Quantum and Topological Systems (CQTS), funded by Tamkeen under the NYUAD Research Institute grant CG008.

\bibliographystyle{ieeetr}
\bibliography{main}

\end{document}